# Beyond Right or Wrong: Towards Redefining Adaptive Learning Indicators in Virtual Learning Environments

**Andréia dos Santos Sachete**
ORCID: https://orcid.org/0000-0003-2226-3322
Federal Institute Farroupilha, Alegrete-RS, Brazil
E-mail: andreia.sachete@iffarroupilha.edu.br

**Alba Valéria de Sant'Anna de Freitas Loiola**
ORCID: https://orcid.org/0000-0003-2418-3393
Federal University of Rio Grande do Sul, Porto Alegre, Brazil
E-mail: alba.portugues@gmail.com

**Fábio Diniz Rossi**
ORCID: https://orcid.org/0000-0002-2450-1024
Federal Institute Farroupilha, Alegrete-RS, Brazil
E-mail: fabio.rossi@iffarroupilha.edu.br

**José Valdeni de Lima**
ORCID: https://orcid.org/0000-0002-7266 -4856
Federal University of Rio Grande do Sul, Porto Alegre, Brazil
E-mail: valdeni@inf.ufrgs.br

**Raquel Salcedo Gomes**
ORCID: https://orcid.org/0000-0001-9497-513X
Federal University of Rio Grande do Sul, Porto Alegre, Brazil
E-mail: raquel.salcedo@ufrgs.br

## ABSTRACT

Student learning development must involve more than just correcting or incorrect questions. However, most adaptive learning methods in Virtual Learning Environments are based on whether the student's response is incorrect or correct. This perspective is limited in assessing the student's learning level, as it does not consider other elements that can be crucial in this process. The objective of this work is to conduct a Systematic Literature Review (SLR) to elucidate which learning indicators influence student learning and which can be implemented in a VLE to assist in adaptive learning. The works selected and filtered by qualitative assessment reveal a comprehensive approach to assessing different aspects of the learning in virtual environments, such as motivation, emotions, physiological responses, brain imaging, and the students' prior knowledge. The discussion of these new indicators allows adaptive technology developers to implement more appropriate solutions to students' realities, resulting in more complete training.

**INTRODUCTION**

Adaptive learning, in the context of virtual learning environments (VLEs), has proven to be a promising approach to optimize the educational process and provide a more effective and personalized learning experience (Maaliw, 2021). With the growing integration of technology in education, these environments have gained prominence as platforms capable of offering a variety of resources and tools that promote and facilitate interactivity, flexibility, and accessibility in teaching (Taylor, Yeung and Bashet, 2021). Adaptive learning seeks to adjust the educational process according to the individual characteristics of each student, recognizing the diversity in pace, of preference and needs. This approach represents an advance over traditional online learning models, which often adopt a unilateral and uniform approach to knowledge transmission. In virtual learning environments, technology can facilitate adaptation, once it allows for the collection and analysis of a vast amount of data on student performance and behavior (Meacham, Pech and Nauck, 2020). Interpreting this data can provide essential insights to individually adjust content, pedagogical approach, and study schedule.

However, most adaptive learning approaches in virtual learning environments rely solely on the student's correct or incorrect response (Sachete *et al.,* 2024). This approach is quite limited in perceiving the student's learning level, as it does not consider other factors that may also be decisive in the learning process (Camillo and Raymundo, 2019; Miquelante *et al.,* 2017). Therefore, for adaptive learning to be effective, it is crucial to determine which learning indicators are relevant and how they can be employed to direct necessary adaptations. Neuroscience delves into the biological processes that underpin learning and memory, exploring how neural networks, synaptic plasticity, and neurotransmitter activity contribute to the acquisition and retention of knowledge (Bin Ibrahim, Benoy and Sajikumar, 2022). By examining them through the lens of neuroscience, researchers can gain insights into how different teaching strategies impact learning processes. For example, studies show that learning strategies that encourage engagement and participation can lead to more robust neural connections, thus enhancing retention and understanding (Munna and Kalam, 2021). This is evidenced by learning metrics, such as higher test scores or faster mastery of complex concepts. Furthermore, neuroscience research has demonstrated the importance of emotions in the learning process (Savolainen, 2019; Shao *et al.*, 2021). It has been found that positive emotional states are associated with more effective formation and retrieval of memories. This connection is often observed in learning metrics, where performance improvement is noticed in educational environments that are free of stress.

Moreover, neuroscientific research on neuroplasticity – the brain's ability to reorganize itself by forming new neural connections throughout life – explains how learning occurs at different ages and stages of development (Parija and Singh, 2023). This can inform the development of appropriate learning metrics that accurately reflect cognitive abilities at various developmental stages. Thus, neuroscience offers a more comprehensive view of the learning process, where the effectiveness of teaching methods is not only assessed through quantitative outcomes or the correctness of responses but is also understood in terms of their impact on the brain's learning mechanisms. This approach can aid educational practices by aligning them more closely with how the brain naturally learns and processes information.

This raises the following question: Which learning indicators influence students' learning, and which can be implemented in a VLE to assist in adaptive learning? This is a complex and multifaceted issue, as learning indicators can encompass a variety of aspects, from traditional ones like performance on tests and activities to more subtle ones like the way students interact with the resources of the virtual environment, the regularity of access, and signs of frustration or engagement on the part of the student.

The contributions to this systematic literature review are listed below:

- A systematic methodology for searching and analyzing articles refers to the indicators used to carry out adaptive learning in virtual learning environments.
- The discussion of the most relevant work in adaptive learning focuses on the indicators used to carry out the adaptation.
- A summarization of the indicators that must be considered by the developers of adaptive learning environments, emphasizing the need for a more holistic view of the student.

Therefore, the importance of conducting a systematic literature review (SLR) to address this gap and answer the previously described question is emphasized. This procedure allows for a deeper understanding of the current state of research, identifying patterns and trends, and can provide a solid foundation for future investigations. In terms of studying adaptive learning in virtual learning environments, an analysis of existing literature in the field of neuroscience enables the exploration of how learning indicators have been addressed in previous studies, the methods used for collecting and analyzing these indicators, and the outcomes achieved in this process. Moreover, the systematic review can also reveal gaps in under-explored learning indicators, such as social interactions among students, the use of multimedia resources, students' self-assessment, and emotional adaptation, which might be under-represented in existing research. Therefore, by identifying the indicators that aid in learning in neuroscience studies, these can be used to propose a virtual learning environment that applies these metrics as a basis to direct and adapt the content and activities according to each student's individuality.

**METHODOLOGY**

A systematic literature review involves a rigorous selection, analysis, and synthesis of relevant existing studies on a specific topic. A vital element of this process is the formulation of a solid protocol, which serves as the basis for planning and executing the review. In this context, we adopt the principles outlined by the Preferred Reporting Items for Systematic Reviews and Meta-Analyses (PRISMA) to develop our systematic review protocol (Page *et al.*, 2021). PRISMA protocol begins with the search for articles using search strings in databases and other relevant sources. For the context of this systematic literature review, we chose to search exclusively in scientific databases. After collecting the results, the articles undergo a filtering process, which includes identifying duplicates, applying inclusion and exclusion criteria, assessing methodological quality, and finally, the remaining articles make up the analysis set that will provide the foundation to answer the research questions addressed by the systematic literature review.

This systematic literature review presents some limitations, which are detailed as follows: The research was conducted solely on articles written in English due to its status as the predominant language in the scientific landscape. Only articles hosted in the digital libraries most frequently mentioned in other systematic literature reviews were selected for analysis: ACM Digital Library, IEEE Digital Library, Science@Direct, and Scopus. The Snowballing technique (Wohlin, 2014), which allows for the incorporation of references from accepted articles aligned with the inclusion and exclusion criteria into the research set, was not employed. Operationally, we used the online platform Parsifal for support. Next, we outline the eligibility criteria, information sources, search strategy, study records, and data synthesis adopted, as recommended in the methods section of the PRISMA checklist. Through the PICOC approach (Population, Intervention,

Comparison, Outcomes, Context), we specify the characteristics of the research to be considered in this review:

- Outcomes: Metrics, indices, or factors that influence learning for students
- Context: Science and neuroscience

We established inclusion and exclusion criteria for the articles to determine the quality of the selection extracted from the databases. The adopted inclusion criteria were:

- Articles written in English.
- Articles published from 2018 onwards (the last 5 years).
- Articles that constitute primary studies.

The chosen exclusion criteria were:

- Articles not pertinent to research.
- Articles of a review or meta-review nature.
- Articles that represent earlier versions of others already considered in this Systematic Literature Review (SLR).

For the search in the information sources, we employed an automatic search sequence, combining keywords and synonyms relevant to two main domains: learning acquisition and neuroscience. The search string was formatted as follows:

(metric OR index OR criterion OR indicat* OR parameter OR factor) AND (learn OR acquire OR obtain OR comprehend OR gather) AND (information OR knowledge OR education) AND neuroscience AND engagement

After collecting all the export files from the search, the materials were integrated into the Parsifal digital data management platform. This tool provides import capabilities and supports the subsequent phases of the process, including the automated detection of duplicate studies and the opportunity to classify each article as accepted or rejected manually.

The consolidation of the exported files established a base containing all potential documents. Initially, an automated process was implemented through Parsifal to remove duplicates. Then, an automated analysis of the titles was conducted, excluding articles that contained the word 'review' to restrict the selection to primary studies. Subsequently, a manual analysis of the journal titles was conducted, as various results from Scopus were not contextualized in the educational field. At this stage, articles from veterinary science and other unrelated disciplines were excluded. Afterward, the titles and abstracts of the remaining articles were read, applying the predefined inclusion and exclusion criteria to refine the results. Articles that met at least one exclusion criterion were not considered. Conversely, an article for inclusion in the final list must meet all the inclusion criteria.

After applying the inclusion and exclusion criteria, we proceeded to read the articles and used a checklist to assess their quality in this final selection. The quality assessment form consisted of five questions with weighted answers: yes (2.0), partially (1.0), not specified (0.0), and no (-0.5). The questions used were as follows:

- Does it discuss learning indicators?
- Does the study present any experiment/application?
- Is the indicator applied/applicable in a Virtual Learning Environment (VLE)?

The maximum possible score was 6.0, obtained from the number of questions and the highest weighted response. Thus, we established the cutoff score at 3.9. Consequently, after reading the content, we excluded articles with scores below this threshold.

**DISCUSION OF RESULTS**

Figure 1 presents a step-by-step description of the phases previously described in the Systematic Literature Review (SLR). The final 16 articles are discussed and summarized at the end of the filtering process.

Figure 1. Systematic review protocol based on PRISMA.

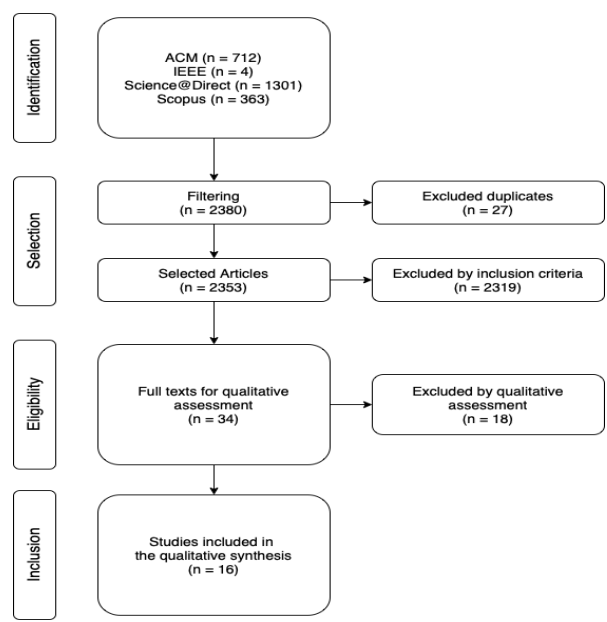

Source: Elaborated by the Authors

Cheng *et al.* (2021) discuss the role of integrating adaptive learning platforms in educational environments, emphasizing the importance of this resource in personalizing students' learning paths based on the individual determination of their knowledge. This ensures that the workload in tasks varies among them. The research problem identified is the challenge of promoting effective engagement of learners in their learning process. It is necessary to ensure that students are motivated and have autonomy in their learning journey, which can lead to better results. The proposed solution involves using an adaptive platform (RealizeIT) that differentiates preparation tasks among students based on their knowledge. This approach ensures learners receive personalized learning experiences, increasing their engagement and understanding. The authors use metrics such as the amount of content covered, task completion, and student progress in content to measure their learning. Furthermore, they highlight the importance of students' engagement efforts in their exam performance, emphasizing the need for a motivating context to improve it. The results indicate that motivated students performed significantly better in multiple-choice and programming questions than unmotivated ones. Moreover, the adaptive learning approach increased student interest, as they demonstrated a stronger sense of belonging and improved confidence levels.

Mercier *et al.* (2020) employed a system dynamics model encompassing affect, cognition, knowledge, performance, context, and external agents. The model was designed to represent the interaction between a student's inherent components (affect, understanding, and knowledge) and external elements (external agents and context), influencing the learning task and, ultimately, the student's performance. The focus of the work is on understanding the temporal dimension of variations related to cognitive load (CL), Cognitive Engagement Index (CEI), Frontal Alpha Asymmetry (an indicator of valence[1], FAA), and Frontal Midline Theta (an indicator of attention[2], FMT). The model seeks to explore the relationships between the components of a learning task, mainly focusing on how affect and cognition are interrelated and how external agents, if present, act on the student's affect and cognition during the learning process. The research involved 72 participants grouped in pairs (player and observer), from whom electroencephalography (EEG) data were collected over two hours, during which one participant was actively playing a serious game designed to teach physics while the other was passively observing. The work employed metrics built from brain imaging (EEG recorded at 1000 Hz) to capture operational task-level activity. These metrics are designed to complement observable behavior records, such as performance traces, verbalizations, and keystrokes/mouse clicks, providing a more comprehensive view of the student's activity and cognitive/affective states during the learning task. The results reveal that each tested metric shows relatively slow cyclic changes, occasionally interspersed with brief episodes of faster variations. The authors also discuss methodological implications and suggest that the typical variation rate observed among variables and participants highlights a potential problem with canceling average data points obtained in longer episodes.

---

[1] Valence refers to two categories of stimuli that can be evaluated: positive (pleasant experience) and negative (unpleasant experience).

[2] FTM is a distinct theta activity of EEG in the frontal midline area that appears during concentrated performance of mental tasks in normal subjects and reflects focused attentional processing.

Tchoubar (2019) investigates the domain of E-learning, highlighting the importance of understanding how people learn using educational technology. The research challenge focused on the need to develop an updated and user-friendly E-learning model, which should integrate conventional learning methods and address new thought processes shaped by digital readiness. For this purpose, Tchoubar (2019) presents a new four-channel E-learning model based on previous cognitive models of multimedia learning and cognitive processing, as it utilizes the model with two primary channels (auditory and visual) and adds kinesthetic and socio-emotional channels. To prove that these last two channels contribute to e-learning, the author described the experiment of their research group, in which they coded and tested a kinesthetic simulation to teach undergraduate students basic electrical circuit designs. Thus, they incorporated the students' spatial competence, which is fundamental for defining the topology of complex circuits, as well as the student's social skills, allowing them to upload their photos to personal profiles and compete virtually with their classmates. Based on this experiment, the author concluded that those with more developed spatial skills performed better and that the kinesthetic and socially interactive interface contributed to the student's learning. The researcher also highlights that, by integrating these new channels with the existing ones, modern learning methods can be built on historical archetypes but elevated through the lens of advanced technology.

Namrata's (2019) research explores digital learning environments and their potential to offer rich and engaging experiences for students. However, it highlights that educational systems often cannot assess students' difficulties in real-time, leading to student frustration. Therefore, it discusses metrics used to measure student learning and their potential applications in educational technology design. These metrics include eye-tracking, thermal imaging, galvanic skin response (GSR), and self-assessment measures. Eye-tracking measures student attention and engagement by tracking their eye movements while interacting with educational materials. Thermal imaging is used to detect changes in facial temperature, which can indicate cognitive load and emotional arousal. GSR measures changes in skin conductance, which can be an index of cognitive load. Self-assessment measures, such as difficulty assessments and learning judgments, are also used to evaluate student learning. The author describes the advantages and limitations of these metrics for assessing student learning and how they can offer significant insights into their cognitive states and levels of engagement. Additionally, the article underlines the need to integrate various metrics to achieve a comprehensive view of the student learning process. It proposes using contactless sensors, such as eye trackers and thermal cameras, to predict students' cognitive load. The study also incorporated a physical linear slider for participants to continuously assess the perceived difficulty of the lessons and a set of sensors that discreetly monitor participants' biometric signals, such as facial expressions, eye movements, and facial temperature. However, the author does not explicitly mention her conclusions but focuses mainly on the methodologies and the potential of multimodal sensors in predicting participants' mental workload, offering design implications for educators to understand the impact of different instructional designs on students' learning gains and experiences.

    Robins (2022) investigates the complexities of cognitive processes, focusing on associative learning. The author describes associative learning as connecting stimuli and responses and as fundamental to the formation of behavior and in the automation of tasks that initially require conscious control. The research problem centers on understanding how different forms of learning, mainly associative and rule-based, interact and influence cognitive performance, especially in tasks requiring automatic and controlled processes. The proposed solution comprehensively explores dual-process theories, emphasizing the role of working memory, cognitive load, and the interaction between associative and rule-based learning mechanisms. The researcher discusses the importance of integrating new information into existing knowledge for effective retention and the role of cognitive load in task execution. The findings presented in the work emphasize the significant role of associative learning in automating tasks over time, resulting in greater efficiency and performance. Furthermore, reinforcement learning algorithms, inspired by associative principles, have also shown performance levels that surpass human capabilities in various domains.

Zhou *et al.* (2018) investigate the domain of cognitive load, exploring its implications in various fields such as human-computer interaction, nursing education, and interactive learning. The researchers describe cognitive load as a multidimensional construct that refers to the momentary load on working memory experienced by a person during the execution of a cognitive task. The study highlights the importance of understanding cognitive load in improving adaptive learning experiences, especially in technologically advanced environments. The central question of the research is to measure and accurately assess cognitive load in diverse scenarios, ensuring that users or learners can effectively manage their tasks, thereby optimizing their performance and learning outcomes. To this end, the authors propose a comprehensive and integrated (multimodal) approach, using physiological sensors, task-evoked pupillary responses, and other innovative methods to accurately measure cognitive load (from questionnaires to complex brain imaging). They also described other techniques, such as galvanic skin response (GSR), digital pen, eye tracker, electrocardiogram (ECG), task completion time, mouse click events, and linguistic patterns. The metrics used to identify cognitive load include the NASA-TLX (Task Load Index), physiological sensors, and the assessment of pupillary responses during mental workload. The results reported by the authors highlight the potential of these metrics in various fields, from detecting driver stress in automotive scenarios to applying the theory of cognitive load in nursing education. Integrating these methods promises a more nuanced understanding of cognitive load, paving the way for enhanced user experiences and learning outcomes. Thus, the authors argue that real-time measurement of cognitive load offers new potential for dynamic support and adaptive system behavior, promising to optimize human-machine interaction. This, in turn, aims to improve human engagement and performance while reducing the pressure on their limited cognitive capacities.

Shao *et al.* (2021) refer to works in the field of affective computing, suggesting that metrics related to the prediction of emotions and personality traits, as well as their relationships, are of fundamental importance. They start from the premise that two primary factors influence an individual's facial reactions during a dyadic interaction: (i) their internal, person-specific cognition and (ii) the non-verbal behaviors exhibited by their conversational partner. Thus, they examine the domain of personality recognition, emphasizing the importance of understanding and modeling person-specific cognitive processes. This study is situated at the intersection of multimedia applications and cognitive science, aiming to provide insights into how individuals' personalities can be inferred from their behavior and interactions. Therefore, the authors use an approach in which they model and represent the cognitive processes of the target subject (the listener) using a unique convolutional neural network (CNN) architecture. The input for this network is the non-verbal audiovisual cues displayed by the conversational partner (the speaker), and the output is the facial reactions of the target subject (the listener). Based on the results, they developed a specific graphical representation of the person, which was used to identify the personality of the target subject. This method aims to enhance the accuracy and robustness of personality recognition systems, capturing the intricate relationships and patterns inherent in human behavior. Furthermore, the results also highlight the potential of the proposed graphical representation method in accurately predicting personality traits from multimedia data, opening a new path for researching socioemotional phenomena (personality, affect, engagement, mental health) from simulations of person-specific cognitive processes, having significant implications for relevant fields, including neuroscience, cognitive and behavioral sciences, among others.

Sweller (2022) builds on evolutionary psychology, exploring its implications for human cognition. The study emphasizes the distinction between biologically primary and biologically secondary information, the latter being culturally essential and acquired through conscious effort. The author addresses the efficiency of different learning methods, contrasting the principle of randomness as genesis (for discovering or inquiring about learning) with the focus on borrowing and reorganization (explicit instruction or learning from others). He shows explicit instruction is more effective than problem-based learning when encountering new information. This claim is supported by the worked-example effect, which shows that studying worked-out examples is more beneficial for students than solving equivalent problems. However, this applies to the acquisition of new and complex information. Therefore, explicit instructions and work examples are recommended over investigative or problem-solving learning methods when beginners acquire new knowledge in a particular field. The author concludes that education should be clear and direct in educational contexts, minimizing unnecessary work memory load and facilitating information transfer to long-term memory. This will ensure that students can more easily assimilate and retain the knowledge acquired during the learning process.

Leisman (2022) discusses the importance of using neuroscientific discoveries to develop innovative teaching methodologies and approaches in educational contexts. He emphasizes the need to understand the brain's functional connectivities and how they are associated with instruction and experience. The author addresses traditional knowledge paradigms that often "medicalize" the process and describes that instead of adopting a binary perspective that simply assesses whether a student has a specific skill, it is more beneficial to focus on the complex processes of the brain that contribute to learning. He suggests that the key lies in understanding the brain's synchrony and how it affects connections between cognitive and motor functions, ultimately influencing the nature of learning. Within this perspective, the author presents that although repetition of material is relevant for effective learning, the literature of cognitive neuroscience highlights the importance of the "spacing effect." This effect refers to students retaining more information when learning sessions are distributed over time rather than concentrated in a single session. Thus, it is beneficial to vary classroom material and reiterate it throughout the semester rather than in a single session or a few days. It is important to note that although we all have similar anatomical brain structures, the neural sets associated with these structures function uniquely for each individual. Therefore, learning tools that can be adapted to meet the specific needs of each student are valuable in the classroom. Leisman (2022) concludes his article by reinforcing that attention should not be directed to binary thinking. Instead, the focus should be on optimized performance, learning strategies, and associative networks, which are more efficiently evaluated through strategic solutions grounded in brain functions.

Sheffler *et al.* (2022) explore the dynamics between cognitive and metacognitive factors, motivational elements, and resources and how these elements affect skill learning at various stages of life. The study focuses on unraveling how these different aspects influence the process of acquiring new skills throughout an individual's life. The proposed methodology offers an in-depth analysis of these factors. Metacognition, for example, is described as a fundamental element, emphasizing "knowing what to learn" and "knowing how to learn," with these two pillars considered essential for the effective acquisition of skills and competencies. Learning new skills varies from childhood to adulthood, with academic motivation often declining during adolescence. However, motivation can be nurtured through the support of teachers, parents, and other social influencers, aiming to keep students engaged, help them overcome challenges, and foster the intrinsic pleasure associated with the learning process. Regarding resources, although there is a diversity of tools available to assist students in acquiring new skills, it's important to note that the distribution and availability of these resources can vary significantly, depending on the individual's life stage and socioeconomic condition. The article explores different metrics and strategies used to assess student learning, including elaboration, which involves connecting new information to prior knowledge, using organizational strategies to systematize their learning process, and monitoring to assess their understanding of newly acquired information. The authors conclude that understanding and leveraging these factors, when considered together, can significantly improve students' learning process. They add that further investigations focused on these three factors, particularly older adult students and how they interact, could enrich our understanding of their impacts on the skill-learning process. These findings, in turn, could serve as a basis for the development of future cognitive interventions, which would be personalized to meet the specific needs of students.

Savolainen (2019) studied the Informational Experience (IE), highlighting sensory and cognitive-affective aspects, as well as the importance of understanding how individuals acquire and interpret information through their senses and cognitive processes. This understanding is essential to relate IE to the broader context of human experience. The central challenge of the research is to define the meaning of IE, understand its connection with human experience, and discern how sensory and cognitive-affective information is acquired and interpreted by various groups of people. The proposed methodology consisted of a conceptual analysis of relevant studies reflecting IE's nature. The goal was to investigate how people receive, seek, remember, and interpret information, aiming for a comprehensive understanding of the sensory and cognitive-affective dimensions of Informational Experience. By treating IE as a lived process, which encompasses the reception, acquisition, and interpretation of sensory and cognitive-affective information, the researcher emphasizes the need to consider both human experience and cognitive interpretation of information to achieve a holistic view of Informational Experience.

Sadafian *et al.* (2019) argue about the evolution of teaching approaches that involve various human senses. This multisensory teaching methodology, also called "dense education," is rooted in the belief that simulating natural learning environments and employing multiple learning modalities (auditory, visual, tactile) can enhance the learning experience. This concept is supported by findings in cognitive neuroscience, which highlight the neural benefits of teaching methods that incorporate multisensory stimuli. The research problem identifies challenges in the consistent application of multisensory teaching techniques. Despite their advantages, more straightforward methods are sometimes only viable due to constraints such as time, inadequate facilities, and specific concepts that may not have a tangible reference in the real world. The suggested approach presents a "dual-continuum model," proposing that while "dense education" (i.e., multisensory teaching) can be beneficial, there are situations where "superficial education" might be more appropriate, especially for adults. This is because adults have complex, experience-dependent cortical connections shaped by various life experiences. The article emphasizes the neurological benefits of multisensory teaching, suggesting that the effectiveness of these methods can be assessed by forming neural networks and fiber connections over time. The results obtained include the demonstration of the cognitive advantages provided by multisensory teaching, as well as the proposal of a dual-continuum model, which can be adapted to meet the needs of different learners and educational contexts.

Gruber *et al.* (2019) explored the concept of curiosity and the neural mechanisms that support it, highlighting the importance of curiosity in forming meaningful learning experiences and its interaction with learning and memory processes. The challenge is to discern how curiosity-driven learning differs from other forms of knowledge and how it can enhance memory retention and retrieval. The aim was to clarify how curiosity-based learning can be optimized. To this end, the article cited studies and experiments that examined the effects of curiosity on memory, postulating that the efficacy of curiosity-driven learning can be measured by its impact on memory retention and recall. The findings or insights discuss the potential of curiosity as a powerful motivator for learning, highlighting the neural advantages of curiosity-driven education and its potential to reinforce memory processes, thus leading to more effective and enduring learning experiences.

Sharma *et al.* (2021) explore the association between the information presented to students and their cognitive processes in the context of digital learning. The research highlights the value of adaptive learning in understanding how students process and engage with digital content. The research problem stems from previous research, which identifies a relationship between the information presented to students, their cognitive load, and changes in attention and focus. The study aims to shift the perspective of analysis from a correlational methodology to a causal approach in the context of multimedia learning. Given this, they propose a solution that uses eye tracking as an effective method to measure cognitive load, attention shifts, and user focus. The study employs a series of measures, including information flow, cognitive load, attention shifts, and user focus size, to derive implications from a causality perspective. The metrics used to assess student learning include eye position (measured in xy coordinates), information flow (quantified as stimulus entropy), cognitive load, user focus size, and attention shift. Data collected from eye tracking and screen recordings were used to calculate these measures, thus providing a comprehensive view of how students interact with and process digital content. The approach to understanding causal relationships in multimedia learning, as opposed to correlational relationships, offers a new perspective on adaptive learning methodologies.

De Mooij *et al.* (2020), focused on online learning environments, specifically those aimed at mathematics, such as the Math Garden platform. These environments are designed to adapt to the individual mathematical abilities of students, thereby maximizing their competencies in this area. The central theme of the research is the potential of adaptive learning in online mathematics education. One of the identified problems is the challenge of online game-based learning environments, which, despite being adaptable, may present interruptions and distractions that affect students' attention, engagement, and concentration. The study aims to analyze how these factors, combined with individual cognitive profiles, impact students' performance in mathematics. The author's approach is to consider not just the students' mathematical abilities but also their cognitive strengths and weaknesses so the learning environment can be tailored to challenge and enhance arithmetic competencies without overloading other areas of their mental capacity. The methodology used to analyze student learning included a computerized arithmetic game, where participants had to choose the correct answer among various options, and eye-tracking technology to understand how students interact with stimuli and how their attention influences problem-solving. The results indicated that adapting online math learning environments to individual cognitive profiles can improve students' learning experience, allowing them to achieve better outcomes in their educational journey.

Chang *et al.* (2019) present a domain of developmental cognitive neuroscience, focusing on children aged eight to ten years at an essential stage for acquiring knowledge. The study explores the neurocognitive mechanisms underpinning learning and the quasi-transfer of problem-solving skills, emphasizing arithmetic, through a five-day math lesson protocol. The study meticulously investigates how learning arithmetic facts through training sessions can result in significant learning, characterized by a clear differentiation of behavioral and brain responses between trained and new problems. The goal is to understand the behavioral, neural, and mnemonic mechanisms that underpin individual peculiarities in children's learning, especially in arithmetic problem-solving competencies, and how these skills are applied in solving new problems after a training period. In the five-day math lesson protocol, children practiced solving addition problems using the separation method for the first three days. They were encouraged to retrieve answers directly from memory in the last two days. An exponential regression model calculates the learning rate, showing significant individual student differences. The metrics used to evaluate student learning included an efficiency score, obtained by dividing accuracy by the average reaction time for correct attempts, and the learning rate, determined using an exponential regression model. The learning curves fit the exponential model well, demonstrating significant individual differences in learning rates. The efficiency score and learning rate were used to measure the progress and development of children over the training days, providing a quantitative indicator of their learning and memory retrieval capabilities. The training process resulted in a significant assimilation of arithmetic facts, characterized by a marked distinction in behavioral and brain responses to previously encountered problems compared to those never seen. Students who exhibited faster learning rates showed more significant performance gains in new but structurally similar questions, indicating successful learning transfer. Furthermore, a multivariate analysis of distance in neural connectivity patterns between trained and new problems showed a significant correlation with the learning rate, suggesting that faster students tend to have a more distinct separation in brain network configurations when dealing with oriented and unknown tasks.

Table 1 summarizes the indicators, metrics, and factors of each of the 16 studies in this SLR. In virtual learning environments, maintaining motivation can be particularly challenging due to the lack of physical presence and direct interaction with teachers, instructors, and peers. Also, the time students spend using the platform is another important metric that provides information about the degree of engagement with the material, learning habits, and how much time they are willing to dedicate to the learning process. Tracking time spent on the platform can also help educators identify which content areas are more challenging or less engaging for students. Other studies have highlighted the importance of sensory modalities – auditory, visual, and kinesthetic – in learning. Additionally, digital platforms can incorporate activities that encourage students to move and interact with the natural environment around them. Although eye-tracking can provide insights into students' engagement, focus, and emotions, it's significant to acknowledge that not all students have access to the necessary technology, such as cameras, which may limit its applicability. Finally, implementing prior knowledge metrics and motivation in virtual learning environments is the foundation for adaptive learning models, which can dynamically adjust content complexity and delivery based on the student's pre-existing understanding and proficiency. This initial assessment optimizes learning pathways, ensuring that learners don't need to review concepts they've already mastered and neither too advanced materials; educators can provide timely feedback and support to assist them.

Table 1. Learning indicators that can contribute to adaptive learning.

| Indicators, Metrics or Factors | Articles |
|---|---|
| Motivation/Engagement | Cheng, Benton e Quinn, 2021<br>Sheffler, Rodriguez, Cheung e Wu, 2022<br>Gruber, Valji e Ranganath, 2019<br>De Mooij, Kirkham, Raijmakers, Van der Maas, e Dumontheil, 2020 |
| Time on the Platform | Cheng, Benton e Quinn, 2021 |
| Emotion analysis | Tchoubar, 2019<br>Zhou, Yu, Chen, Wang e Arshad, 2018<br>Shao, Song, Jaiswal, Shen, Valstar, Gunes, 2021<br>Savolainen, 2019 |
| Multisensory (visual, auditory, kinesthetic) | Mercier, Whissell-Turner, Paradis, and Avaca, 2020<br>Tchoubar, 2019<br>Namrata, 2019<br>Zhou, Yu, Chen, Wang e Arshad, 2018<br>Sadafian, Pishghadam, Moghimi, 2019 |
| Prior Knowledge | Robins, 2022<br>Sweller, 2022<br>Leisman, 2022<br>Sheffler, Rodriguez, Cheung e Wu, 2022<br>Chang, Rosenberg-Lee, Qin, Menon, 2019 |
| Other physiological responses (brain imaging; galvanic skin responses, eye tracker) | Mercier, Whissell-Turner, Paradis, and Avaca, 2020<br>Namrata, 2019<br>Zhou, Yu, Chen, Wang e Arshad, 2018<br>Sadafian, Pishghadam, Moghimi, 2019<br>Sharma, Mangaroska, Berkel, Giannakos, e Kostakos, 2021<br>De Mooij, Kirkham, Raijmakers, Van der Maas, e Dumontheil, 2020 |

Source: Elaborated by the Authors

**CONCLUSIONS**

The studies analyzed in this research highlighted the importance of various metrics in measuring learning. Motivation, platform usage time, emotions, multisensory, physiological responses, and prior recognition can provide a comprehensive, data-based approach to understanding student performance and engagement, offering more effective and personalized learning experiences. These indicators are particularly relevant in the context of adaptive learning models, which rely on data and analytics to tailor learning content to students' individual needs and preferences. By leveraging these metrics, adaptive learning models can provide targeted support and feedback to help students overcome challenges and succeed in their learning process.

However, there is still a long way to go, as most learning platforms do not rely on more than one metric to assess student progress. Many only use limited data, such as time spent on the platform or assessment scores, which does not provide a complete picture of student performance and engagement. This approach can lead to less personalized and less effective learning experiences, as important factors such as motivation, emotions, or physiological responses need to be considered, which are important for a more comprehensive understanding of the learning process. Therefore, for adaptive learning models to reach their full potential, platforms must begin to integrate a wider range of metrics into their analytics.